\documentclass[nofootinbib,preprint]{revtex4}%
\usepackage{hyperref}
\usepackage{amsmath}
\usepackage{amsfonts}
\usepackage{amssymb}
\usepackage{graphicx}%
\begin{document}
\title{Notes on High Energy Limit of Bosonic Closed String Scattering Amplitudes}
\author{Chuan-Tsung Chan}
\email{ctchan@phys.cts.nthu.edu.tw}
\affiliation{Physics Division, National Center for Theoretical Sciences, Hsinchu, Taiwan, R.O.C.}
\author{Jen-Chi Lee}
\email{jcclee@cc.nctu.edu.tw}
\affiliation{Department of Electrophysics, National Chiao-Tung University, Hsinchu, Taiwan, R.O.C.}
\author{Yi Yang}
\email{yiyang@mail.nctu.edu.tw}
\affiliation{Department of Electrophysics, National Chiao-Tung University, Hsinchu, Taiwan, R.O.C.}
\date{\today}

\begin{abstract}
We study bosonic closed string scattering amplitudes in the high-energy limit.
We find that the methods of decoupling of high-energy zero-norm states and the
high-energy Virasoro constraints, which were adopted in the previous works to
calculate the ratios among high-energy open string scattering amplitudes of
different string states, persist for the case of closed string. However, we
clarify the previous saddle-point calculation for high-energy open string
scattering amplitudes and claim that only $(t,u)$ channel of the amplitudes is
suitable for saddle-point calculation. We then discuss three evidences to show
that saddle-point calculation for high-energy closed string scattering
amplitudes is not reliable. By using the relation of tree-level closed and
open string scattering amplitudes of Kawai, Lewellen and Tye (KLT), we
calculate the high-energy closed string scattering amplitudes for
\textit{arbitrary} mass levels. For the case of high-energy closed string
four-tachyon amplitude, our result differs from the previous one of Gross and
Mende, which is NOT consistent with KLT formula, by an oscillating factor.

\end{abstract}
\maketitle

\section{Introduction}

Recently high-energy, fixed-angle behavior of string scattering amplitudes
\cite{GM, Gross, GrossManes} was intensively reinvestigated
\cite{ChanLee1,ChanLee2, CHL,CHLTY,PRL,paperB,susy}. The motivation was to
uncover the long-sought hidden stringy space-time symmetry. An important new
ingredient of this approach is the zero-norm states (ZNS)
\cite{ZNS1,ZNS3,ZNS2} in the old covariant first quantized (OCFQ) string
spectrum. One utilizes the decoupling of zero-norm states to obtain relations
among scattering amplitudes. An infinite number of linear relations among
high-energy scattering amplitudes of different string states were derived.
Moreover, these linear relations can be used to fix the proportionality
constants among high-energy scattering amplitudes of different string states
at each fixed mass level algebraically. Thus there is only one independent
component of high-energy scattering amplitude at each fixed mass level. On the
other hand, a saddle-point method was also developed to calculate the general
formula of tree-level high-energy scattering amplitudes of four arbitrary
string states to verify the ratios among the high-energy scattering amplitudes
of different string states calculated by the above algebraic methods.
Moreover, these high-energy scattering amplitudes can be expressed in terms of
high-energy four tachyon  scattering amplitude as conjectured by Gross in 1988
\cite{Gross}. However, all the above calculations were focused only on the
case of open string theory.

In this paper, we generalize the calculations to high-energy closed string
scattering amplitudes. We find that the methods of decoupling of high-energy
zero-norm states and the high-energy Virasoro constraints, which were adopted
in the previous works to calculate the ratios among high-energy open string
scattering amplitudes of different string states, persist for the case of
closed string. The result is simply the tensor product of two pieces of open
string ratios of high-energy scattering amplitudes. However, we clarify the
previous saddle-point calculation for high-energy open string scattering
amplitudes and claim that only $(t,u)$ channel of the amplitudes is suitable
for saddle-point calculation. We then discuss three evidences to show that
saddle-point calculation for high-energy closed string scattering amplitudes
is not reliable. By using the relation of tree-level closed and open string
scattering amplitudes of Kawai, Lewellen and Tye (KLT) \cite{KLT}, we
calculate the tree-level high-energy closed string scattering amplitudes for
\textit{arbitrary} mass levels. For the case of high-energy closed string
four-tachyon amplitude, our result differs from the previous one of Gross and
Mende \cite{GM}, which is NOT consistent with KLT formula, by an oscillating
factor. This means that the high-energy closed string amplitudes \textit{do
not factorize} into product of two high-energy open string amplitudes in
contrast to the conventional wisdom \cite{GM,Moore}.

\section{Decoupling of Zero Norm States}

In this section, we calculate the ratios among high-energy closed string
scattering amplitudes of different string states by the decoupling of
high-energy closed string ZNS. Since the calculation is similar to that of
open string, we will, for simplicity, work on the first massive level
$M^{2}=8(n-1)=8$ $\left(  n=2\right)  $ only. At this mass level, the
corresponding open string Ward identities are ($M^{2}=2$ for open string,
$\alpha_{\text{closed}}^{\prime}=4\alpha_{\text{open}}^{\prime}=2$)
\cite{ChanLee3}
\begin{align}
k_{\mu}\theta_{\nu}\mathcal{T}^{\mu\nu}+\theta_{\mu}\mathcal{T}^{\mu}  &
=0,\\
\left(  \frac{3}{2}k_{\mu}k_{\nu}+\frac{1}{2}\eta_{\mu\nu}\right)
\mathcal{T}^{\mu\nu}+\frac{5}{2}k_{\mu}\mathcal{T}^{\mu}  &  =0,
\end{align}
where $\theta_{\nu}$ is a transverse vector. In Eqs.(1) and (2), we have
chosen, say, the second vertex $V_{2}(k_{2})$ to be the vertex operators
constructed from zero-norm states and $k_{\mu}\equiv k_{2\mu}$. The other
three vertices can be any string states. Note that Eq.(1) is the type I Ward
identity while Eq.(2) is the type II Ward identity which is valid only at
$D=26$. The high-energy limits of Eqs.(1) and (2) were calculated to be
\begin{align}
M\mathcal{T}_{TP}^{3\rightarrow1}+\mathcal{T}_{T}^{1}  &  =0,\\
M\mathcal{T}_{LL}^{4\rightarrow2}+\mathcal{T}_{L}^{2}  &  =0,\\
3M^{2}\mathcal{T}_{LL}^{4\rightarrow2}+\mathcal{T}_{TT}^{2}+5M\mathcal{T}%
_{L}^{2}  &  =0.
\end{align}
In the above equations, we have defined the following orthonormal polarization
vectors for the second string vertex $V_{2}(k_{2})$
\begin{align}
e_{P}  &  =\frac{1}{M}(E_{2},\mathrm{k}_{2},0)=\frac{k_{2}}{M},\\
e_{L}  &  =\frac{1}{M}(\mathrm{k}_{2},E_{2},0),\\
e_{T}  &  =(0,0,1)
\end{align}
in the center-of-mass (\emph{c.m.}) frame contained in the plane of
scattering. We have also denoted the naive power counting for orders in energy
\cite{ChanLee1,ChanLee2} in the superscript of each amplitude according to the
following rules, $e_{L}\cdot k\sim E^{2},e_{T}\cdot k\sim E^{1}$. Note that
since $\mathcal{T}_{TP}^{1}$ is of subleading order in energy, in general
$\mathcal{T}_{TP}^{1}\neq\mathcal{T}_{TL}^{1}$. A simple calculation of
Eqs.(3)-(5) shows that \cite{ChanLee3}
\begin{equation}
\mathcal{T}_{TP}^{1}:\mathcal{T}_{T}^{1}=1:-\sqrt{2}=1:-M.
\end{equation}%
\begin{equation}
\mathcal{T}_{TT}^{2}:\mathcal{T}_{LL}^{2}:\mathcal{T}_{L}^{2}\text{ }=\text{
}4:1:-\sqrt{2}=2M^{2}:1:-M.
\end{equation}
It is interesting to see that, in addition to the leading order amplitudes in
Eq.(10), the subleading order amplitudes in Eq.(9) are also proportional to
each other. This does not seem to happen at higher mass level.

We are now back to the closed string calculation. The OCFQ closed string
spectrum at this mass level are
$(\raisebox{0.06in}{\fbox{\rule[0.04cm]{0.04cm}{0cm}}}\raisebox{0.06in}{\fbox{\rule[0.04cm]{0.04cm}{0cm}}}+\raisebox{0.06in}{\fbox{\rule[0.04cm]{0.04cm}{0cm}}}+\bullet
)\otimes
(\raisebox{0.06in}{\fbox{\rule[0.04cm]{0.04cm}{0cm}}}\raisebox{0.06in}{\fbox{\rule[0.04cm]{0.04cm}{0cm}}}+\raisebox{0.06in}{\fbox{\rule[0.04cm]{0.04cm}{0cm}}}+\bullet
)^{^{\prime}}.$ In addition to the spin-four positive-norm state
$\raisebox{0.06in}{\fbox{\rule[0.04cm]{0.04cm}{0cm}}}\raisebox{0.06in}{\fbox{\rule[0.04cm]{0.04cm}{0cm}}}\otimes
\raisebox{0.06in}{\fbox{\rule[0.04cm]{0.04cm}{0cm}}}\raisebox{0.06in}{\fbox{\rule[0.04cm]{0.04cm}{0cm}}}^{\prime
}$, one has 8 ZNS, each of which gives a Ward identity. In the high-energy
limit, we have $\theta^{\mu\nu}=e_{L}^{\mu}e_{L}^{\nu}-e_{T}^{\mu}e_{T}^{\nu}$
or $\theta^{\mu\nu}=e_{L}^{\mu}e_{T}^{\nu}+e_{T}^{\mu}e_{L}^{\nu}$,
$\theta^{\mu}=e_{L}^{\mu}$ or $e_{T}^{\mu}$ and one replace $\eta_{\mu\nu}$ by
$e_{T}^{\mu}e_{T}^{\nu}$. In the following, we list only high-energy Ward
identities which relate amplitudes with even-energy power in the high-energy
expansion :

\noindent\bigskip1.
$\raisebox{0.06in}{\fbox{\rule[0.04cm]{0.04cm}{0cm}}}\raisebox{0.06in}{\fbox{\rule[0.04cm]{0.04cm}{0cm}}}\otimes
\raisebox{0.06in}{\fbox{\rule[0.04cm]{0.04cm}{0cm}}}^{^{\prime}}:$%
\begin{align}
M(\mathcal{T}_{LL,LL}-\mathcal{T}_{TT,LL})+\mathcal{T}_{LL,L}-\mathcal{T}%
_{TT,L}  &  =0,\\
M\mathcal{T}_{LT,PT}+\mathcal{T}_{LT,T}  &  =0.
\end{align}
2.
$\raisebox{0.06in}{\fbox{\rule[0.04cm]{0.04cm}{0cm}}}\raisebox{0.06in}{\fbox{\rule[0.04cm]{0.04cm}{0cm}}}\otimes
\bullet^{^{\prime}}:$%
\begin{equation}
3M^{2}(\mathcal{T}_{LL,LL}-\mathcal{T}_{TT,LL})+(\mathcal{T}_{LL,TT}%
-\mathcal{T}_{TT,TT})+5M(\mathcal{T}_{LL,L}-\mathcal{T}_{TT,L})=0.
\end{equation}
3. $\raisebox{0.06in}{\fbox{\rule[0.04cm]{0.04cm}{0cm}}}\otimes
\raisebox{0.06in}{\fbox{\rule[0.04cm]{0.04cm}{0cm}}}\raisebox{0.06in}{\fbox{\rule[0.04cm]{0.04cm}{0cm}}}^{^{\prime
}}:$%
\begin{align}
M(\mathcal{T}_{LL,LL}-\mathcal{T}_{LL,TT})+\mathcal{T}_{L,LL}-\mathcal{T}%
_{L,TT}  &  =0,\\
M\mathcal{T}_{PT,LT}+\mathcal{T}_{T,LT}  &  =0.
\end{align}
4. $\raisebox{0.06in}{\fbox{\rule[0.04cm]{0.04cm}{0cm}}}\otimes
\raisebox{0.06in}{\fbox{\rule[0.04cm]{0.04cm}{0cm}}}^{^{\prime}}:$%
\begin{align}
M^{2}\mathcal{T}_{LL,LL}+M\mathcal{T}_{LL,L}+M\mathcal{T}_{L,LL}%
+\mathcal{T}_{L,L}  &  =0,\\
M^{2}\mathcal{T}_{PT,PT}+M\mathcal{T}_{PT,T}+M\mathcal{T}_{T,PT}%
+\mathcal{T}_{T,T}  &  =0.
\end{align}
5. $\raisebox{0.06in}{\fbox{\rule[0.04cm]{0.04cm}{0cm}}}\otimes\bullet
^{^{\prime}}:$%
\begin{equation}
3M^{3}\mathcal{T}_{LL,LL}+M\mathcal{T}_{LL,TT}+5M^{2}\mathcal{T}_{LL,L}%
+3M^{2}\mathcal{T}_{L,LL}+\mathcal{T}_{L,TT}+5M^{2}\mathcal{T}_{L,L}=0.
\end{equation}
6. $\bullet\otimes
\raisebox{0.06in}{\fbox{\rule[0.04cm]{0.04cm}{0cm}}}\raisebox{0.06in}{\fbox{\rule[0.04cm]{0.04cm}{0cm}}}^{\prime
}:$%
\begin{equation}
3M^{2}(\mathcal{T}_{LL,LL}-\mathcal{T}_{LL,TT})+(\mathcal{T}_{TT,LL}%
-\mathcal{T}_{TT,TT})+5M(\mathcal{T}_{L,LL}-\mathcal{T}_{L,TT})=0.
\end{equation}
7. $\bullet\otimes
\raisebox{0.06in}{\fbox{\rule[0.04cm]{0.04cm}{0cm}}}^{^{\prime}}:$%
\begin{equation}
3M^{3}\mathcal{T}_{LL,LL}+M\mathcal{T}_{TT,LL}+5M^{2}\mathcal{T}_{L,LL}%
+3M^{2}\mathcal{T}_{LL,L}+\mathcal{T}_{TT,L}+5M^{2}\mathcal{T}_{L,L}=0.
\end{equation}
8. $\bullet\otimes\bullet^{^{\prime}}:$%
\begin{align}
&  9M^{4}\mathcal{T}_{LL,LL}+3M^{2}\mathcal{T}_{LL,TT}+3M^{2}\mathcal{T}%
_{TT,LL}+15M^{3}\mathcal{T}_{LL,L}\nonumber\\
&  +15M^{3}\mathcal{T}_{L,LL}+5M\mathcal{T}_{TT,L}+5M\mathcal{T}%
_{L,TT}+25M^{2}\mathcal{T}_{L,L}+\mathcal{T}_{TT,TT}=0.
\end{align}
Those Ward identities which relate amplitudes with odd-energy power in the
high-energy expansion are omitted as they are subleading order in energy. The
mass $M$ in Eqs.(11) to (21) should now be interpreted as the closed string
mass $M^{2}=8$. Eqs.(12),(15) and (17) are subleading order amplitudes, and
one can then solve the other 8 Eqs. to give the ratios%
\begin{align}
&  \mathcal{T}_{TT,TT} :\mathcal{T}_{TT,LL}:\mathcal{T}_{LL,TT}:\mathcal{T}%
_{LL,LL}:\mathcal{T}_{TT,L}:\mathcal{T}_{L,TT}:\mathcal{T}_{LL,L}%
:\mathcal{T}_{L,LL}:\mathcal{T}_{L,L}\nonumber\\
=  &  1:\frac{1}{2M^{2}}:\frac{1}{2M^{2}}:\frac{1}{4M^{4}}:-\frac{1}%
{2M}:-\frac{1}{2M}:-\frac{1}{4M^{3}}:-\frac{1}{4M^{3}}:\frac{1}{4M^{2}}.
\end{align}
Eq.(22) is exactly the tensor product of two pieces of open string ratios
calculated in Eq.(10).

\section{Virasoro Constraints}

We consider the mass level $M^{2}=8$ $\left(  n=2\right)  $. The most general
state is%
\begin{align}
\left\vert 2\right\rangle  &  =\left\{  \frac{1}{2!}%
\begin{tabular}
[c]{|c|c|}\hline
$\mu_{1}^{1}$ & $\mu_{2}^{1}$\\\hline
\end{tabular}
\ \alpha_{-1}^{\mu_{1}^{1}}\alpha_{-1}^{\mu_{2}^{1}}+\frac{1}{2}%
\begin{tabular}
[c]{|c|}\hline
$\mu_{1}^{2}$\\\hline
\end{tabular}
\ \alpha_{-2}^{\mu_{1}^{2}}\right\}  \otimes\left\{  \frac{1}{2!}%
\begin{tabular}
[c]{|c|c|}\hline
$\tilde{\mu}_{1}^{1}$ & $\tilde{\mu}_{2}^{1}$\\\hline
\end{tabular}
\ \tilde{\alpha}_{-1}^{\tilde{\mu}_{1}^{1}}\tilde{\alpha}_{-1}^{\tilde{\mu
}_{2}^{1}}+\frac{1}{2}%
\begin{tabular}
[c]{|c|}\hline
$\tilde{\mu}_{1}^{2}$\\\hline
\end{tabular}
\ \tilde{\alpha}_{-2}^{\tilde{\mu}_{1}^{2}}\right\}  \left\vert
0,k\right\rangle \nonumber\\
&  =\frac{1}{4}\left\{
\begin{tabular}
[c]{|c|c|}\hline
$\mu_{1}^{1}$ & $\mu_{2}^{1}$\\\hline
\end{tabular}
\ \alpha_{-1}^{\mu_{1}^{1}}\alpha_{-1}^{\mu_{2}^{1}}+%
\begin{tabular}
[c]{|c|}\hline
$\mu_{1}^{2}$\\\hline
\end{tabular}
\ \alpha_{-2}^{\mu_{1}^{2}}\right\}  \otimes\left\{
\begin{tabular}
[c]{|c|c|}\hline
$\tilde{\mu}_{1}^{1}$ & $\tilde{\mu}_{2}^{1}$\\\hline
\end{tabular}
\ \tilde{\alpha}_{-1}^{\tilde{\mu}_{1}^{1}}\tilde{\alpha}_{-1}^{\tilde{\mu
}_{2}^{1}}+%
\begin{tabular}
[c]{|c|}\hline
$\tilde{\mu}_{1}^{2}$\\\hline
\end{tabular}
\ \tilde{\alpha}_{-2}^{\tilde{\mu}_{1}^{2}}\right\}  \left\vert
0,k\right\rangle .
\end{align}
The Virasoro constraints are%
\begin{subequations}
\begin{align}
L_{1}\left\vert 2\right\rangle  &  \sim\left\{  k^{\mu_{1}^{1}}%
\begin{tabular}
[c]{|c|c|}\hline
$\mu_{1}^{1}$ & $\mu_{2}^{1}$\\\hline
\end{tabular}
\ \alpha_{-1}^{\mu_{2}^{1}}+%
\begin{tabular}
[c]{|c|}\hline
$\mu_{1}^{2}$\\\hline
\end{tabular}
\ \alpha_{-1}^{\mu_{1}^{2}}\right\}  \otimes\left\{
\begin{tabular}
[c]{|c|c|}\hline
$\tilde{\mu}_{1}^{1}$ & $\tilde{\mu}_{2}^{1}$\\\hline
\end{tabular}
\ \tilde{\alpha}_{-1}^{\tilde{\mu}_{1}^{1}}\tilde{\alpha}_{-1}^{\tilde{\mu
}_{2}^{1}}+%
\begin{tabular}
[c]{|c|}\hline
$\tilde{\mu}_{1}^{2}$\\\hline
\end{tabular}
\ \tilde{\alpha}_{-2}^{\tilde{\mu}_{1}^{2}}\right\}  =0,\\
\tilde{L}_{1}\left\vert 2\right\rangle  &  \sim\left\{
\begin{tabular}
[c]{|c|c|}\hline
$\mu_{1}^{1}$ & $\mu_{2}^{1}$\\\hline
\end{tabular}
\ \alpha_{-1}^{\mu_{1}^{1}}\alpha_{-1}^{\mu_{2}^{1}}+%
\begin{tabular}
[c]{|c|}\hline
$\mu_{1}^{2}$\\\hline
\end{tabular}
\ \alpha_{-2}^{\mu_{1}^{2}}\right\}  \otimes\left\{  k^{\mu_{1}^{1}}%
\begin{tabular}
[c]{|c|c|}\hline
$\tilde{\mu}_{1}^{1}$ & $\tilde{\mu}_{2}^{1}$\\\hline
\end{tabular}
\ \tilde{\alpha}_{-1}^{\tilde{\mu}_{2}^{1}}+%
\begin{tabular}
[c]{|c|}\hline
$\tilde{\mu}_{1}^{2}$\\\hline
\end{tabular}
\ \tilde{\alpha}_{-1}^{\tilde{\mu}_{1}^{2}}\right\}  =0,\\
L_{2}\left\vert 2\right\rangle  &  \sim\left\{
\begin{tabular}
[c]{|c|c|}\hline
$\mu_{1}^{1}$ & $\mu_{2}^{1}$\\\hline
\end{tabular}
\ \eta^{\mu_{1}^{1}\mu_{2}^{1}}+2k^{\mu_{1}^{2}}%
\begin{tabular}
[c]{|c|}\hline
$\mu_{1}^{2}$\\\hline
\end{tabular}
\ \right\}  \otimes\left\{
\begin{tabular}
[c]{|c|c|}\hline
$\tilde{\mu}_{1}^{1}$ & $\tilde{\mu}_{2}^{1}$\\\hline
\end{tabular}
\ \tilde{\alpha}_{-1}^{\tilde{\mu}_{1}^{1}}\tilde{\alpha}_{-1}^{\tilde{\mu
}_{2}^{1}}+%
\begin{tabular}
[c]{|c|}\hline
$\tilde{\mu}_{1}^{2}$\\\hline
\end{tabular}
\ \tilde{\alpha}_{-2}^{\tilde{\mu}_{1}^{2}}\right\}  =0,\\
\tilde{L}_{2}\left\vert 2\right\rangle  &  \sim\left\{
\begin{tabular}
[c]{|c|c|}\hline
$\mu_{1}^{1}$ & $\mu_{2}^{1}$\\\hline
\end{tabular}
\ \alpha_{-1}^{\mu_{1}^{1}}\alpha_{-1}^{\mu_{2}^{1}}+%
\begin{tabular}
[c]{|c|}\hline
$\mu_{1}^{2}$\\\hline
\end{tabular}
\ \alpha_{-2}^{\mu_{1}^{2}}\right\}  \otimes\left\{
\begin{tabular}
[c]{|c|c|}\hline
$\tilde{\mu}_{1}^{1}$ & $\tilde{\mu}_{2}^{1}$\\\hline
\end{tabular}
\ \eta^{\tilde{\mu}_{12}^{1\tilde{\mu}1}}+2k^{\tilde{\mu}_{1}^{2}}%
\begin{tabular}
[c]{|c|}\hline
$\tilde{\mu}_{1}^{2}$\\\hline
\end{tabular}
\ \right\}  =0.
\end{align}
Taking the high-energy limit in the above equations by letting $\left(
\mu_{i},\nu_{i}\right)  \rightarrow\left(  L,T\right)  $, and%
\end{subequations}
\begin{equation}
k^{\mu_{i}}\rightarrow Me^{L}\text{, }\eta^{\mu_{1}\mu_{2}}\rightarrow
e^{T}e^{T},
\end{equation}
we obtain
\begin{subequations}
\begin{align}
\left\{  M%
\begin{tabular}
[c]{|c|c|}\hline
$L$ & $\mu$\\\hline
\end{tabular}
\ +%
\begin{tabular}
[c]{|c|}\hline
$\mu$\\\hline
\end{tabular}
\ \right\}  \alpha_{-1}^{\mu}\otimes\left\{
\begin{tabular}
[c]{|c|c|}\hline
$\tilde{\mu}_{1}^{1}$ & $\tilde{\mu}_{2}^{1}$\\\hline
\end{tabular}
\ \tilde{\alpha}_{-1}^{\tilde{\mu}_{1}^{1}}\tilde{\alpha}_{-1}^{\tilde{\mu
}_{2}^{1}}+%
\begin{tabular}
[c]{|c|}\hline
$\tilde{\mu}_{1}^{2}$\\\hline
\end{tabular}
\ \tilde{\alpha}_{-2}^{\tilde{\mu}_{1}^{2}}\right\}   &  =0,\\
\left\{
\begin{tabular}
[c]{|c|c|}\hline
$\mu_{1}^{1}$ & $\mu_{2}^{1}$\\\hline
\end{tabular}
\ \alpha_{-1}^{\mu_{1}^{1}}\alpha_{-1}^{\mu_{2}^{1}}+%
\begin{tabular}
[c]{|c|}\hline
$\mu_{1}^{2}$\\\hline
\end{tabular}
\ \alpha_{-2}^{\mu_{1}^{2}}\right\}  \otimes\left\{  M%
\begin{tabular}
[c]{|c|c|}\hline
$L$ & $\tilde{\mu}$\\\hline
\end{tabular}
\ +%
\begin{tabular}
[c]{|c|}\hline
$\tilde{\mu}$\\\hline
\end{tabular}
\ \right\}  \tilde{\alpha}_{-1}^{\tilde{\mu}} &  =0,\\
\left\{
\begin{tabular}
[c]{|c|c|}\hline
$T$ & $T$\\\hline
\end{tabular}
\ +2M%
\begin{tabular}
[c]{|c|}\hline
$L$\\\hline
\end{tabular}
\ \right\}  \otimes\left\{
\begin{tabular}
[c]{|c|c|}\hline
$\tilde{\mu}_{1}^{1}$ & $\tilde{\mu}_{2}^{1}$\\\hline
\end{tabular}
\ \tilde{\alpha}_{-1}^{\tilde{\mu}_{1}^{1}}\tilde{\alpha}_{-1}^{\tilde{\mu
}_{2}^{1}}+%
\begin{tabular}
[c]{|c|}\hline
$\tilde{\mu}_{1}^{2}$\\\hline
\end{tabular}
\ \tilde{\alpha}_{-2}^{\tilde{\mu}_{1}^{2}}\right\}   &  =0,\\
\left\{
\begin{tabular}
[c]{|c|c|}\hline
$\mu_{1}^{1}$ & $\mu_{2}^{1}$\\\hline
\end{tabular}
\ \alpha_{-1}^{\mu_{1}^{1}}\alpha_{-1}^{\mu_{2}^{1}}+%
\begin{tabular}
[c]{|c|}\hline
$\mu_{1}^{2}$\\\hline
\end{tabular}
\ \alpha_{-2}^{\mu_{1}^{2}}\right\}  \otimes\left\{
\begin{tabular}
[c]{|c|c|}\hline
$T$ & $T$\\\hline
\end{tabular}
\ +2M%
\begin{tabular}
[c]{|c|}\hline
$L$\\\hline
\end{tabular}
\ \right\}   &  =0,
\end{align}
which lead to the following equations%
\end{subequations}
\begin{subequations}
\begin{align}
\left\{  M%
\begin{tabular}
[c]{|c|c|}\hline
$L$ & $\mu$\\\hline
\end{tabular}
\ +%
\begin{tabular}
[c]{|c|}\hline
$\mu$\\\hline
\end{tabular}
\ \right\}  \otimes%
\begin{tabular}
[c]{|c|c|}\hline
$\tilde{\mu}_{1}^{1}$ & $\tilde{\mu}_{2}^{1}$\\\hline
\end{tabular}
\  &  =0,\\
\left\{  M%
\begin{tabular}
[c]{|c|c|}\hline
$L$ & $\mu$\\\hline
\end{tabular}
\ +%
\begin{tabular}
[c]{|c|}\hline
$\mu$\\\hline
\end{tabular}
\ \right\}  \otimes%
\begin{tabular}
[c]{|c|}\hline
$\tilde{\mu}_{1}^{2}$\\\hline
\end{tabular}
\  &  =0,\\%
\begin{tabular}
[c]{|c|c|}\hline
$\mu_{1}^{1}$ & $\mu_{2}^{1}$\\\hline
\end{tabular}
\ \otimes\left\{  M%
\begin{tabular}
[c]{|c|c|}\hline
$L$ & $\tilde{\mu}$\\\hline
\end{tabular}
\ +%
\begin{tabular}
[c]{|c|}\hline
$\tilde{\mu}$\\\hline
\end{tabular}
\ \right\}   &  =0,\\%
\begin{tabular}
[c]{|c|}\hline
$\mu_{1}^{2}$\\\hline
\end{tabular}
\ \otimes\left\{  M%
\begin{tabular}
[c]{|c|c|}\hline
$L$ & $\tilde{\mu}$\\\hline
\end{tabular}
\ +%
\begin{tabular}
[c]{|c|}\hline
$\tilde{\mu}$\\\hline
\end{tabular}
\ \right\}   &  =0,\\
\left\{
\begin{tabular}
[c]{|c|c|}\hline
$T$ & $T$\\\hline
\end{tabular}
\ +2M%
\begin{tabular}
[c]{|c|}\hline
$L$\\\hline
\end{tabular}
\ \right\}  \otimes%
\begin{tabular}
[c]{|c|c|}\hline
$\tilde{\mu}_{1}^{1}$ & $\tilde{\mu}_{2}^{1}$\\\hline
\end{tabular}
\  &  =0,\\
\left\{
\begin{tabular}
[c]{|c|c|}\hline
$T$ & $T$\\\hline
\end{tabular}
\ +2M%
\begin{tabular}
[c]{|c|}\hline
$L$\\\hline
\end{tabular}
\ \right\}  \otimes%
\begin{tabular}
[c]{|c|}\hline
$\tilde{\mu}_{1}^{2}$\\\hline
\end{tabular}
\  &  =0,\\%
\begin{tabular}
[c]{|c|c|}\hline
$\mu_{1}^{1}$ & $\mu_{2}^{1}$\\\hline
\end{tabular}
\ \otimes\left\{
\begin{tabular}
[c]{|c|c|}\hline
$T$ & $T$\\\hline
\end{tabular}
\ +2M%
\begin{tabular}
[c]{|c|}\hline
$L$\\\hline
\end{tabular}
\ \right\}   &  =0,\\%
\begin{tabular}
[c]{|c|}\hline
$\mu_{1}^{2}$\\\hline
\end{tabular}
\ \otimes\left\{
\begin{tabular}
[c]{|c|c|}\hline
$T$ & $T$\\\hline
\end{tabular}
\ +2M%
\begin{tabular}
[c]{|c|}\hline
$L$\\\hline
\end{tabular}
\ \right\}   &  =0.
\end{align}
The remaining indices $\mu,\tilde{\mu}$ in the above equations can be set to
be $T$ or $L$, and we obtain%
\end{subequations}
\begin{subequations}
\begin{align}
M%
\begin{tabular}
[c]{|c|c|}\hline
$L$ & $L$\\\hline
\end{tabular}
\ \otimes%
\begin{tabular}
[c]{|c|c|}\hline
$L$ & $L$\\\hline
\end{tabular}
\ +%
\begin{tabular}
[c]{|c|}\hline
$L$\\\hline
\end{tabular}
\ \otimes%
\begin{tabular}
[c]{|c|c|}\hline
$L$ & $L$\\\hline
\end{tabular}
\  &  =0,\label{1}\\
M%
\begin{tabular}
[c]{|c|c|}\hline
$L$ & $L$\\\hline
\end{tabular}
\ \otimes%
\begin{tabular}
[c]{|c|c|}\hline
$T$ & $T$\\\hline
\end{tabular}
\ +%
\begin{tabular}
[c]{|c|}\hline
$L$\\\hline
\end{tabular}
\ \otimes%
\begin{tabular}
[c]{|c|c|}\hline
$T$ & $T$\\\hline
\end{tabular}
\  &  =0,\label{2}\\
M%
\begin{tabular}
[c]{|c|c|}\hline
$T$ & $L$\\\hline
\end{tabular}
\ \otimes%
\begin{tabular}
[c]{|c|c|}\hline
$T$ & $L$\\\hline
\end{tabular}
\ +%
\begin{tabular}
[c]{|c|}\hline
$T$\\\hline
\end{tabular}
\ \otimes%
\begin{tabular}
[c]{|c|c|}\hline
$T$ & $L$\\\hline
\end{tabular}
\  &  =0,\label{3}%
\end{align}%
\end{subequations}
\begin{subequations}
\begin{align}
M%
\begin{tabular}
[c]{|c|c|}\hline
$L$ & $L$\\\hline
\end{tabular}
\ \otimes%
\begin{tabular}
[c]{|c|}\hline
$L$\\\hline
\end{tabular}
\ +%
\begin{tabular}
[c]{|c|}\hline
$L$\\\hline
\end{tabular}
\ \otimes%
\begin{tabular}
[c]{|c|}\hline
$L$\\\hline
\end{tabular}
\  &  =0,\label{4}\\
M%
\begin{tabular}
[c]{|c|c|}\hline
$T$ & $L$\\\hline
\end{tabular}
\ \otimes%
\begin{tabular}
[c]{|c|}\hline
$T$\\\hline
\end{tabular}
\ +%
\begin{tabular}
[c]{|c|}\hline
$T$\\\hline
\end{tabular}
\ \otimes%
\begin{tabular}
[c]{|c|}\hline
$T$\\\hline
\end{tabular}
\  &  =0,\label{5}%
\end{align}%
\end{subequations}
\begin{subequations}
\begin{align}
M%
\begin{tabular}
[c]{|c|c|}\hline
$L$ & $L$\\\hline
\end{tabular}
\ \otimes%
\begin{tabular}
[c]{|c|c|}\hline
$L$ & $L$\\\hline
\end{tabular}
\ +%
\begin{tabular}
[c]{|c|c|}\hline
$L$ & $L$\\\hline
\end{tabular}
\ \otimes%
\begin{tabular}
[c]{|c|}\hline
$L$\\\hline
\end{tabular}
\  &  =0,\label{6}\\
M%
\begin{tabular}
[c]{|c|c|}\hline
$T$ & $T$\\\hline
\end{tabular}
\ \otimes%
\begin{tabular}
[c]{|c|c|}\hline
$L$ & $L$\\\hline
\end{tabular}
\ +%
\begin{tabular}
[c]{|c|c|}\hline
$T$ & $T$\\\hline
\end{tabular}
\ \otimes%
\begin{tabular}
[c]{|c|}\hline
$L$\\\hline
\end{tabular}
\  &  =0,\label{7}\\
M%
\begin{tabular}
[c]{|c|c|}\hline
$T$ & $L$\\\hline
\end{tabular}
\ \otimes%
\begin{tabular}
[c]{|c|c|}\hline
$T$ & $L$\\\hline
\end{tabular}
\ +%
\begin{tabular}
[c]{|c|c|}\hline
$T$ & $L$\\\hline
\end{tabular}
\ \otimes%
\begin{tabular}
[c]{|c|}\hline
$T$\\\hline
\end{tabular}
\  &  =0,\label{8}%
\end{align}%
\end{subequations}
\begin{subequations}
\begin{align}
M%
\begin{tabular}
[c]{|c|}\hline
$L$\\\hline
\end{tabular}
\ \otimes%
\begin{tabular}
[c]{|c|c|}\hline
$L$ & $L$\\\hline
\end{tabular}
\ +%
\begin{tabular}
[c]{|c|}\hline
$L$\\\hline
\end{tabular}
\ \otimes%
\begin{tabular}
[c]{|c|}\hline
$L$\\\hline
\end{tabular}
\  &  =0,\label{9}\\
M%
\begin{tabular}
[c]{|c|}\hline
$T$\\\hline
\end{tabular}
\ \otimes%
\begin{tabular}
[c]{|c|c|}\hline
$T$ & $L$\\\hline
\end{tabular}
\ +%
\begin{tabular}
[c]{|c|}\hline
$T$\\\hline
\end{tabular}
\ \otimes%
\begin{tabular}
[c]{|c|}\hline
$T$\\\hline
\end{tabular}
\  &  =0,\label{10}%
\end{align}%
\end{subequations}
\begin{subequations}
\begin{align}%
\begin{tabular}
[c]{|c|c|}\hline
$T$ & $T$\\\hline
\end{tabular}
\ \otimes%
\begin{tabular}
[c]{|c|c|}\hline
$L$ & $L$\\\hline
\end{tabular}
\ +2M%
\begin{tabular}
[c]{|c|}\hline
$L$\\\hline
\end{tabular}
\ \otimes%
\begin{tabular}
[c]{|c|c|}\hline
$L$ & $L$\\\hline
\end{tabular}
\  &  =0,\label{11}\\%
\begin{tabular}
[c]{|c|c|}\hline
$T$ & $T$\\\hline
\end{tabular}
\ \otimes%
\begin{tabular}
[c]{|c|c|}\hline
$T$ & $T$\\\hline
\end{tabular}
\ +2M%
\begin{tabular}
[c]{|c|}\hline
$L$\\\hline
\end{tabular}
\ \otimes%
\begin{tabular}
[c]{|c|c|}\hline
$T$ & $T$\\\hline
\end{tabular}
\  &  =0,\label{12}%
\end{align}%
\end{subequations}
\begin{equation}%
\begin{tabular}
[c]{|c|c|}\hline
$T$ & $T$\\\hline
\end{tabular}
\ \otimes%
\begin{tabular}
[c]{|c|}\hline
$L$\\\hline
\end{tabular}
\ +2M%
\begin{tabular}
[c]{|c|}\hline
$L$\\\hline
\end{tabular}
\ \otimes%
\begin{tabular}
[c]{|c|}\hline
$L$\\\hline
\end{tabular}
\ =0,\label{13}%
\end{equation}%
\begin{subequations}
\begin{align}%
\begin{tabular}
[c]{|c|c|}\hline
$L$ & $L$\\\hline
\end{tabular}
\ \otimes%
\begin{tabular}
[c]{|c|c|}\hline
$T$ & $T$\\\hline
\end{tabular}
\ +2M%
\begin{tabular}
[c]{|c|c|}\hline
$L$ & $L$\\\hline
\end{tabular}
\ \otimes%
\begin{tabular}
[c]{|c|}\hline
$L$\\\hline
\end{tabular}
\  &  =0,\label{14}\\%
\begin{tabular}
[c]{|c|c|}\hline
$T$ & $T$\\\hline
\end{tabular}
\ \otimes%
\begin{tabular}
[c]{|c|c|}\hline
$T$ & $T$\\\hline
\end{tabular}
\ +2M%
\begin{tabular}
[c]{|c|c|}\hline
$T$ & $T$\\\hline
\end{tabular}
\ \otimes%
\begin{tabular}
[c]{|c|}\hline
$L$\\\hline
\end{tabular}
\  &  =0,\label{15}%
\end{align}%
\end{subequations}
\begin{equation}%
\begin{tabular}
[c]{|c|}\hline
$L$\\\hline
\end{tabular}
\ \otimes%
\begin{tabular}
[c]{|c|c|}\hline
$T$ & $T$\\\hline
\end{tabular}
\ +2M%
\begin{tabular}
[c]{|c|}\hline
$L$\\\hline
\end{tabular}
\ \otimes%
\begin{tabular}
[c]{|c|}\hline
$L$\\\hline
\end{tabular}
\ =0.\label{16}%
\end{equation}
Since the transverse component of the highest spin state $\alpha_{-1}%
^{T}\cdots\alpha_{-1}^{T}\otimes\tilde{\alpha}_{-1}^{T}\cdots\tilde{\alpha
}_{-1}^{T}$ at each fixed mass level gives the leading order scattering
amplitude, there should have even number of $T$ at each fixed mass level. Thus
Eqs. (\ref{3}), (\ref{5}), (\ref{8}) and (\ref{10}) are subleading order in
energy and are therefore irrelevant. Set $%
\begin{tabular}
[c]{|c|c|}\hline
$T$ & $T$\\\hline
\end{tabular}
\ \otimes%
\begin{tabular}
[c]{|c|c|}\hline
$T$ & $T$\\\hline
\end{tabular}
\ =1$, we can solve the ratios from the remaining equations. The final result is

\begin{center}
\noindent%
\begin{tabular}
[c]{|c|c|}\hline
$\epsilon_{TT,TT}$ & $1$\\\hline
$\epsilon_{TT,LL}=\epsilon_{LL,TT}$ & $1/\left(  2M^{2}\right)  $\\\hline
$\epsilon_{LL,LL}$ & $1/\left(  4M^{4}\right)  $\\\hline
$\epsilon_{TT,L}=\epsilon_{L,TT}$ & $-1/\left(  2M\right)  $\\\hline
$\epsilon_{LL,L}=\epsilon_{L,LL}$ & $-1/\left(  4M^{3}\right)  $\\\hline
$\epsilon_{L,L}$ & $1/\left(  4M^{2}\right)  $\\\hline
\end{tabular}

\end{center}

\noindent which is exactly the tensor product of two pieces of open string
ratios. This result is consistent with Eq.(22) from the decoupling of
high-energy zero-norm state in section II.

\section{Saddle Point Calculation}

In this section, we calculate the tree-level high-energy closed string
scattering amplitudes for arbitrary mass levels. We first review the
calculation of high-energy open string scattering amplitude. The $(s,t)$
channel scattering amplitude with $V_{2}=\alpha_{-1}^{\mu_{1}}\alpha_{-1}%
^{\mu_{2}}..\alpha_{-1}^{\mu_{n}}\mid0,k>$, the highest spin state at mass
level $M^{2}$ $=2(n-1),$ and three tachyons $V_{1,3,4}$ is \cite{CHL}%
\begin{equation}
\mathcal{T}_{n;st}^{\mu_{1}\mu_{2}\cdot\cdot\mu_{n}}=\overset{n}%
{\underset{l=0}{\sum}}(-)^{l}\binom{n}{l}B\left(  -\frac{s}{2}-1+l,-\frac
{t}{2}-1+n-l\right)  k_{1}^{(\mu_{1}}..k_{1}^{\mu_{n-l}}k_{3}^{\mu_{n-l+1}%
}..k_{3}^{\mu_{n})},\label{B}%
\end{equation}
where $B(u,v)=\int_{0}^{1}dxx^{u-1}(1-x)^{v-1}$ is the Euler beta function. It
is now easy to calculate the general high-energy scattering amplitude at the
$M^{2}$ $=2(n-1)$ level
\begin{equation}
\mathcal{T}_{n}^{TTT\cdot\cdot}(s,t)\simeq\lbrack-2E^{3}\sin\phi_{c.m.}%
]^{n}\mathcal{T}_{n}(s,t)\label{T}%
\end{equation}
where $\mathcal{T}_{n}(s,t)$ is the high energy limit of $\frac{\Gamma
(-\frac{s}{2}-1)\Gamma(-\frac{t}{2}-1)}{\Gamma(\frac{u}{2}+2)}$ with
$s+t+u=2n-8$, and was previously \cite{ChanLee1,CHL} miscalculated to be%
\begin{align}
\mathcal{\tilde{T}}_{n;st} &  \simeq\sqrt{\pi}(-1)^{n-1}2^{-n}E^{-1-2n}\left(
\sin\frac{\phi_{c.m.}}{2}\right)  ^{-3}\left(  \cos\frac{\phi_{c.m.}}%
{2}\right)  ^{5-2n}\nonumber\\
&  \times\exp\left[  -\frac{s\ln s+t\ln t-(s+t)\ln(s+t)}{2}\right]
.\label{st}%
\end{align}
One can now generalize this result to multi-tensors. The $(s,t)$ channel of
open string high-energy scattering amplitude at mass level $(n_{1},n_{2}%
,n_{3},n_{4})$ was calculated to be \cite{ChanLee1,CHL}%

\begin{equation}
\mathcal{T}_{n_{1}n_{2}n_{3}n_{4};st}^{T^{1}\cdot\cdot T^{2}\cdot\cdot
T^{3}\cdot\cdot T^{4}\cdot\cdot}=[-2E^{3}\sin\phi_{c.m.}]^{\Sigma n_{i}%
}\mathcal{T}_{\Sigma n_{i}}(s,t).\label{AM}%
\end{equation}
In the above calculations, the scattering angle $\phi_{c.m.}$ in the center of
mass frame is defined to be the angle between $\overrightarrow{k}_{1}$ and
$\overrightarrow{k}_{3}$. $s=-(k_{1}+k_{2})^{2}$, $t=-(k_{2}+k_{3})^{2}$ and
$u=-(k_{1}+k_{3})^{2}$ are the Mandelstam variables. $M_{i}^{2}=2(n_{i}-1)$
with $n_{i}$ the mass level of the $i$th vertex. $T^{i}$ in Eq.(\ref{AM}) is
the transverse polarization of the $i$th vertex defined in Eq.(8). All other
4-point functions at mass level $(n_{1},n_{2},n_{3},n_{4})$ were shown to be
proportional to Eq.(\ref{AM}).

The corresponding $(t,u)$ channel scattering amplitudes of Eqs.(\ref{T}) and
(\ref{AM}) can be obtained by replacing $(s,t)$ in Eq.(\ref{st}) by $(t,u)$
\begin{align}
\mathcal{T}_{n}(t,u)  &  \simeq\sqrt{\pi}(-1)^{n-1}2^{-n}E^{-1-2n}\left(
\sin\frac{\phi_{c.m.}}{2}\right)  ^{-3}\left(  \cos\frac{\phi_{c.m.}}%
{2}\right)  ^{5-2n}\nonumber\\
&  \times\exp\left[  -\frac{t\ln t+u\ln u-(t+u)\ln(t+u)}{2}\right]  .
\label{tu}%
\end{align}

We now claim that only $(t,u)$ channel of the amplitude, Eq.(\ref{tu}), is
suitable for saddle-point calculation. The previous saddle-point calculation
for the $(s,t)$ channel amplitude, Eq.(\ref{st}), in the high-energy expansion
is misleading. The corrected high-energy calculation of the $(s,t)$ channel
amplitude will be given in Eq.(\ref{corr}). The reason is as following. When
calculating Eq.(\ref{T}) from Eq.(\ref{B}), one calculates the high-energy
limit of%
\begin{equation}
\frac{\Gamma(-\frac{s}{2}-1)\Gamma(-\frac{t}{2}-1)}{\Gamma(\frac{u}{2}%
+2)},s+t+u=2n-8,
\end{equation}
in Eq.(\ref{B}) by expanding the $\Gamma$ function with the Stirling formula%
\begin{equation}
\Gamma\left(  x\right)  \sim\sqrt{2\pi}x^{x-1/2}e^{-x}.
\end{equation}
However, the above expansion is not suitable for negative real $x$ as there
are poles for $\Gamma\left(  x\right)  $ at $x=-n$, negative integers.
Unfortunately, our high-energy limit
\begin{subequations}
\label{high energy limit}%
\begin{align}
s &  \sim4E^{2}\gg0,\\
t &  \sim-4E^{2}\sin^{2}\left(  \frac{\phi_{c.m.}}{2}\right)  \ll0,\\
u &  \sim-4E^{2}\cos^{2}\left(  \frac{\phi_{c.m.}}{2}\right)  \ll0,
\end{align}
contains this dangerous situation in the $(s,t)$ channel calculation of
Eq.(\ref{st}). On the other hand, the corresponding high-energy expansion of
$(t,u)$ channel scattering amplitude in Eq.(\ref{tu}) is well defined. Another
evidence for this point is the following. When one uses the saddle point
method to calculate the high-energy open string scattering amplitudes in the
$(s,t)$ channel, the saddle-point we identified was \cite{CHL,CHLTY,PRL}%
\end{subequations}
\begin{equation}
x_{0}=\frac{s}{s+t}=\frac{1}{1-\sin^{2}\left(  \phi/2\right)  }>1,
\end{equation}
which is out of the integration range $\left(  0,1\right)  $. Therefore, we
can not trust the saddle point calculation for the $(s,t)$ channel scattering
amplitude. On the other hand, the corresponding saddle-point calculation for
the $(t,u)$ channel scattering amplitude is safe since the saddle-point
$x_{0}$ is within the integration range $\left(  1,\infty\right)  $. This
subtle situation becomes crucial and relevant when one tries to calculate the
high-energy closed string scatterings amplitude and compare them with the open
string ones.

We now discuss the high-energy closed string scattering amplitudes. There
exists a celebrated formula by Kawai, Lewellen and Tye (KLT), which expresses
the relation between tree amplitudes of closed and open string $(\alpha
_{\text{closed}}^{\prime}=4\alpha_{\text{open}}^{\prime}=2)$
\begin{equation}
A_{\text{closed}}^{\left(  4\right)  }\left(  s,t,u\right)  =\sin\left(  \pi
k_{2}\cdot k_{3}\right)  A_{\text{open}}^{\left(  4\right)  }\left(
s,t\right)  \bar{A}_{\text{open}}^{\left(  4\right)  }\left(  t,u\right)
.\label{KLT}%
\end{equation}
To calculate the high-energy closed string scattering amplitudes, one
encounters the difficulty of calculation of high-energy open string amplitude
in the $(s,t)$ channel discussed above. To avoid this difficulty, we can use
the well known formula
\begin{equation}
\Gamma\left(  x\right)  =\frac{\pi}{\sin\left(  \pi x\right)  \Gamma\left(
1-x\right)  }%
\end{equation}
to calculate the large negative $x$ expansion of the $\Gamma$ function. We
first discuss the high-energy four-tachyon scattering amplitude which already
existed in the literature. We can express the open string $(s,t)$ channel
amplitude in terms of the $(t,u)$ channel amplitude,
\begin{align}
A_{\text{open}}^{\left(  4\text{-tachyon}\right)  }\left(  s,t\right)   &
=\frac{\Gamma\left(  -\frac{s}{2}-1\right)  \Gamma\left(  -\frac{t}%
{2}-1\right)  }{\Gamma\left(  \frac{u}{2}+2\right)  }\nonumber\\
&  =\frac{\sin\left(  \pi u/2\right)  }{\sin\left(  \pi s/2\right)  }%
\frac{\Gamma\left(  -\frac{t}{2}-1\right)  \Gamma\left(  -\frac{u}%
{2}-1\right)  }{\Gamma\left(  \frac{s}{2}+2\right)  }\nonumber\\
&  \equiv\frac{\sin\left(  \pi u/2\right)  }{\sin\left(  \pi s/2\right)
}A_{\text{open}}^{\left(  4\text{-tachyon}\right)  }\left(  t,u\right)
,\label{Tach}%
\end{align}
which we know how to calculate the high-energy limit. \ Note that for the
four-tachyon case,$\ \bar{A}_{\text{open}}^{\left(  4\right)  }\left(
t,u\right)  =A_{\text{open}}^{\left(  4\right)  }\left(  t,u\right)  $ in
Eq.(\ref{KLT}). The KLT formula, Eq.(\ref{KLT}), can then be used to express
the closed string four-tachyon scattering amplitude in terms of that of open
string in the $(t,u)$ channel
\begin{equation}
A_{\text{closed}}^{\left(  4\text{-tachyon}\right)  }\left(  s,t,u\right)
=\frac{\sin\left(  \pi t/2\right)  \sin\left(  \pi u/2\right)  }{\sin\left(
\pi s/2\right)  }A_{\text{open}}^{\left(  4\text{-tachyon}\right)  }\left(
t,u\right)  A_{\text{open}}^{\left(  4\text{-tachyon}\right)  }\left(
t,u\right)  .
\end{equation}
The high-energy limit of open string four-tachyon amplitude in the $(t,u)$
channel can be easily calculated to be%
\begin{equation}
A_{\text{open}}^{(4-\text{tachyon})}\left(  t,u\right)  \simeq(stu)^{-\frac
{3}{2}}\exp\left(  -\frac{s\ln s+t\ln t+u\ln u}{2}\right)  ,\label{open}%
\end{equation}
which gives the corresponding amplitude in the $(s,t)$ channel%
\begin{equation}
A_{\text{open}}^{(4-\text{tachyon})}\left(  s,t\right)  \simeq\frac
{\sin\left(  \pi u/2\right)  }{\sin\left(  \pi s/2\right)  }(stu)^{-\frac
{3}{2}}\exp\left(  -\frac{s\ln s+t\ln t+u\ln u}{2}\right)  .\label{sttachyon}%
\end{equation}
The high-energy limit of closed string four-tachyon scattering amplitude can
then be calculated, through the KLT formula, to be%
\begin{equation}
A_{\text{closed}}^{(4-\text{tachyon})}\left(  s,t,u\right)  \simeq\frac
{\sin\left(  \pi t/2\right)  \sin\left(  \pi u/2\right)  }{\sin\left(  \pi
s/2\right)  }(stu)^{-3}\exp\left(  -\frac{s\ln s+t\ln t+u\ln u}{4}\right)
.\label{new}%
\end{equation}
The exponential factor in Eq.(\ref{open}) was first discussed by Veneziano
\cite{Veneziano}. Our result for the high-energy closed string four-tachyon
amplitude in Eq.(\ref{new}) differs from the one calculated in the literature
\cite{GM} by an oscillating factor $\frac{\sin\left(  \pi t/2\right)
\sin\left(  \pi u/2\right)  }{\sin\left(  \pi s/2\right)  }$\cite{GSW}.
\textit{We stress here that our results for Eqs.(\ref{open}), (\ref{sttachyon}%
) and (\ref{new}) are consistent with the KLT formula, while the previous
calculation in \cite{GM} is NOT.}

One might try to use the saddle-point method to calculate the high-energy
closed string scattering amplitude. The closed string four-tachyon scattering
amplitude is%
\begin{align}
A_{\text{closed}}^{(4-\text{tachyon})}\left(  s,t,u\right)   &  =\int
dxdy\exp\left(  \frac{k_{1}\cdot k_{2}}{2}\ln\left\vert z\right\vert
+\frac{k_{2}\cdot k_{3}}{2}\ln\left\vert 1-z\right\vert \right)  \nonumber\\
&  =\int dxdy(x^{2}+y^{2})^{-2}[(1-x)^{2}+y^{2}]^{-2}\exp\left\{  -\frac{s}%
{8}\ln(x^{2}+y^{2})-\frac{t}{8}\ln[(1-x)^{2}+y^{2}]\right\}  \nonumber\\
&  \equiv\int dxdy(x^{2}+y^{2})^{-2}[(1-x)^{2}+y^{2}]^{-2}\exp\left[
-Kf(x,y)\right]
\end{align}
where $K=\frac{s}{8}$ and $f(x,y)=\ln(x^{2}+y^{2})-\tau\ln[(1-x)^{2}+y^{2}]$
with $\tau=-\frac{t}{s}$. One can then calculate the "saddle-point" of
$\ f(x,y)$ to be%
\begin{equation}
\nabla f(x,y)\mid_{x_{0}=\frac{1}{1-\tau},y_{0}=0}=0.
\end{equation}
The high-energy limit of the closed string four-tachyon scattering amplitude
is then calculated to be%
\begin{equation}
A_{\text{closed}}^{(4-\text{tachyon})}\left(  s,t,u\right)  \simeq\frac{2\pi
}{K\sqrt{\det\frac{\partial^{2}f(x_{0},y_{0})}{\partial x\partial y}}}%
\exp[-Kf(x_{0},y_{0})]\simeq(stu)^{-3}\exp\left(  -\frac{s\ln s+t\ln t+u\ln
u}{4}\right)  ,
\end{equation}
which is consistent with the previous one calculated in the literature
\cite{GM}, but is different from our result in Eq.(\ref{new}). However, one
notes that%
\begin{equation}
\frac{\partial^{2}f(x_{0},y_{0})}{\partial x^{2}}=\frac{2(1-\tau)^{3}}{\tau
}=-\frac{\partial^{2}f(x_{0},y_{0})}{\partial y^{2}},\frac{\partial^{2}%
f(x_{0},y_{0})}{\partial x\partial y}=0,
\end{equation}
which means that $(x_{0},y_{0})$ is NOT the local minimum of $f(x,y)$, and one
should not trust this saddle-point calculation. This is the third evidence to
see that there is no clear definition of saddle-point in the calculation of
the high-energy open string scattering amplitude in the $(s,t)$ channel, and
thus the invalid saddle-point calculation of high-energy closed string
scattering amplitude.

Finally we calculate the high-energy closed string scattering amplitudes for
arbitrary mass levels. The $(t,u)$ channel open string scattering amplitude
with $V_{2}=\alpha_{-1}^{\mu_{1}}\alpha_{-1}^{\mu_{2}}..\alpha_{-1}^{\mu_{n}%
}\mid0,k>$, the highest spin state at mass level $M^{2}$ $=2(n-1)$, and three
tachyons $V_{1,3,4}$ can be calculated to be%
\begin{equation}
\mathcal{T}_{n;tu}^{\mu_{1}\mu_{2}\cdot\cdot\mu_{n}}=\overset{n}%
{\underset{l=0}{\sum}}\binom{n}{l}B\left(  -\frac{t}{2}+n-l-1,-\frac{u}%
{2}-1\right)  k_{1}^{(\mu_{1}}..k_{1}^{\mu_{n-l}}k_{3}^{\mu_{n-l+1}}%
..k_{3}^{\mu_{n})}.\label{ntu}%
\end{equation}
In calculating Eq.(\ref{ntu}), we have used the Mobius transformation
$y=\frac{x-1}{x}$ to change the integration region from $\left(
1,\infty\right)  $ to $\left(  0,1\right)  $. One notes that Eq.(\ref{ntu}) is
NOT the same as Eq.(\ref{B}) with $(s,t)$ replaced by $(t,u)$, as one would
have expected from the four-tachyon case discussed in the paragraph after
Eq.(\ref{KLT}) \ In the high-energy limit, one easily sees that%
\begin{equation}
\mathcal{T}_{n}(s,t)\simeq(-)^{n}\frac{\sin\left(  \pi u/2\right)  }%
{\sin\left(  \pi s/2\right)  }\mathcal{T}_{n}(t,u),\label{corr}%
\end{equation}
which is the generalization of Eq.(\ref{Tach}) to arbitrary mass levels.
Eq.(\ref{corr}) is the correction of Eqs.(\ref{T}) and (\ref{st}) as claimed
in the paragraph after Eq.(\ref{tu}). The $(s,t)$ channel of high-energy open
string scattering amplitudes at mass level $(n_{1},n_{2},n_{3},n_{4})$ can
then be written as, apart from an overall constant,%
\begin{align}
A_{\text{open}}^{\left(  4\right)  }\left(  s,t\right)   &  \simeq(-)^{\Sigma
n_{i}}\frac{\sin\left(  \pi u/2\right)  }{\sin\left(  \pi s/2\right)
}[-2E^{3}\sin\phi_{c.m.}]^{\Sigma n_{i}}\mathcal{T}_{\Sigma n_{i}%
}(t,u)\nonumber\\
&  \simeq(-)^{\Sigma n_{i}}\frac{\sin\left(  \pi u/2\right)  }{\sin\left(  \pi
s/2\right)  }(stu)^{\frac{\Sigma n_{i}-3}{2}}\exp\left(  -\frac{s\ln s+t\ln
t+u\ln u}{2}\right)  .\label{openst}%
\end{align}
Finally the total high-energy open string scattering amplitude is the sum of
$\left(  s,t\right)  $, $\left(  t,u\right)  $ and $\left(  u,s\right)  $
channel amplitudes, and can be calculated to be%
\begin{equation}
A_{\text{open}}^{\left(  4\right)  }\simeq(-)^{\Sigma n_{i}}\frac{\sin\left(
\pi s/2\right)  +\sin\left(  \pi t/2\right)  +\sin\left(  \pi u/2\right)
}{\sin\left(  \pi s/2\right)  }(stu)^{\frac{\Sigma n_{i}-3}{2}}\exp\left(
-\frac{s\ln s+t\ln t+u\ln u}{2}\right)  .\label{openall}%
\end{equation}
By using Eqs.(\ref{KLT}) and (\ref{corr}), the high-energy closed string
scattering amplitude at mass level $(n_{1},n_{2},n_{3},n_{4})$ is calculated
to be, apart from an overall constant,%
\begin{align}
A_{\text{closed}}^{\left(  4\right)  }\left(  s,t,u\right)   &  \simeq
(-)^{\Sigma n_{i}}\frac{\sin\left(  \pi t/2\right)  \sin\left(  \pi
u/2\right)  }{\sin\left(  \pi s/2\right)  }[-2E^{3}\sin\phi_{c.m.}]^{2\Sigma
n_{i}}\mathcal{T}_{\Sigma n_{i}}(t,u)^{2}\nonumber\\
&  \simeq(-)^{\Sigma n_{i}}\frac{\sin\left(  \pi t/2\right)  \sin\left(  \pi
u/2\right)  }{\sin\left(  \pi s/2\right)  }(stu)^{\Sigma n_{i}-3}\exp\left(
-\frac{s\ln s+t\ln t+u\ln u}{4}\right)  ,\label{SA}%
\end{align}
where $\mathcal{T}_{\Sigma n_{i}}(t,u)$ is given by Eq.(\ref{tu}). For the
case of four-tachyon scattering amplitude at mass level $(0,0,0,0)$,
Eq.(\ref{SA}) reduces to Eq.(\ref{new}). All other high-energy closed string
scattering amplitudes at mass level $(n_{1},n_{2},n_{3},n_{4})$ are
proportional to Eq.(\ref{SA}). The proportionality constants are the tensor
product of two pieces of open string ratios.

\section{Conclusion}

In conclusion, we have used the methods of decoupling of high-energy zero-norm
states and the high-energy Virasoro constraints to calculate the ratios among
high-energy closed string scattering amplitudes of different string states.
The result is exactly the tensor product of two pieces of open string ratios
calculated before. However, we clarify the previous saddle-point calculation
for high-energy open string scattering amplitudes and show that only $(t,u)$
channel of the amplitudes is suitable for saddle-point calculation. We also
discuss three evidences, Eqs.(43),(44) and (55), to show that saddle-point
calculation for high-energy closed string scattering amplitudes is not
reliable. Instead of using saddle-point calculation adopted before, we then
propose to use the formula of Kawai, Lewellen and Tye (KLT) to calculate the
high-energy closed string scattering amplitudes for \textit{arbitrary} mass
levels. For the case of high-energy closed string four-tachyon amplitude, our
result differs from the previous one of Gross and Mende, which is NOT
consistent with KLT formula, by an oscillating factor. The oscillating
prefactors in Eqs.(\ref{openall}) and (\ref{SA}) imply the existence of
infinitely many zeros and poles in the string scattering amplitudes even in
the high-energy limit. Physically, the presence of poles simply reflects the
fact that there are infinite number of resonances in the string spectrum
\cite{GSW}, and the presence of zeros reflects the coherence of string scattering.

\section{Acknowledgments}

This work is supported in part by the National Science Council and National
Center for Theoretical Science, Taiwan, R.O.C.

\end{document}